\documentclass[aps,prl,twocolumn,showpacs,floatfix]{revtex4}

\usepackage{amsmath}

\usepackage{graphicx}

\usepackage{bm}

\begin{document}

\title{Wave packet dynamics in chains with delayed electronic nonlinear response}

\author{F.~A.~B.~F.~de Moura,Iram Gleria, I.~F.~dos Santos, and M.~L.~Lyra}
\affiliation{Instituto de  F\'{\i}sica, Universidade Federal de
Alagoas, Macei\'{o} AL 57072-970, Brazil}

\begin{abstract}
 
We  study the dynamics of one electron wave packet in a chain with a non-adiabatic electron-phonon interaction. The electron-phonon coupling is taken into account in the time-dependent Schr\"odinger equation by a delayed cubic nonlinearity.   In the limit of an adiabatic coupling, the self-trapping phenomenon occurs when the nonlinearity parameter exceeds a critical value of the order of the band width. We show that  a weaker nonlinearity is required to produce self-trapping in the regime of short delay times. However, this trend is reversed for slow nonlinear responses, resulting in a reentrant phase-diagram. In slowly responding media, self-trapping only takes place for very strong nonlinearities.

\end{abstract}

\maketitle

The study of the physical mechanisms involved in transport phenomena taking place in nonlinear systems is a fundamental issue in solid state physics. Concerning the electronic transport, nonlinearity arises from the interaction between electrons
and lattice vibrations~\cite{Johansson,kundu,Xiong1,ivanchenko,flach,skokos,capone}. In this context, the discrete nonlinear Schr\"odinger equation (DNLSE) effectively describes the influence of lattice vibrations on the electron dynamics. The most important property associated with the DNLSE is the self-trapping which occurs when the nonlinearity
parameter exceeds a critical value of the order of the band width~\cite{Johansson,kundu,Xiong1}. In this regime, an initially localized electronic wave packet does not spread continuously over the lattice. Therefore, the probability of finding the electron at its
initial site remains finite in the long-time limit.  

In low-dimensional systems, the effect of nonlinearity  seems to be dominant over the role played  by disorder~\cite{prl1,prl2,prl3,prl4}. Recently, the spreading of an initially localized wave packet in two nonlinear chains with disorder was studied~\cite{prl1}. Considering a discrete nonlinear
Schr\"odinger and quartic Klein-Gordon equations with disorder, it was proved that the second moment  and the participation number of the wave packet do not diverge simultaneously~\cite{prl1}. The spreading of an initially localized wave packet in a one-dimensional discrete nonlinear Schr\"odinger lattice with disorder was also recently studied~\cite{prl2}. It was observed that the Anderson localization is destroyed and a subdiffusive dynamics takes place above a certain critical nonlinearity strength.~\cite{prl2} Moreover, analytical and numerical calculations for a reduced Fermi-Pasta-Ulam chain demonstrated that energy localization does not require more than one conserved quantity~\cite{prl3}. 

From the experimental point of view, the interplay between disorder and nonlinearity  was investigated in Ref.~\cite{prl4}. The evolution of linear and nonlinear waves in coupled optical waveguides patterned on an AlGaAs substrate
 were directly measured. Nonlinear perturbations enhance localization of linear waves while induce delocalization of the nonlinear ones~\cite{prl4}. In the presence of disorder, a transition from ballistic wave
packet expansion to exponential localization was observed. Within a more general scope, the study of wave propagation in nonlinear systems is a very interesting problem with several applications in many branches of physics. For example, by using a model with diagonal and off-diagonal electron-phonon-like terms, the role played by the degree of integrability and the resulting extreme event statistic were recently studied~\cite{sea}. The main  motivation was  to explain  the sudden appearance  sea waves of extremely large amplitude in relatively calm seas. 

Usually, the nonlinearity produced by the electron-phonon interaction is assumed to be instantaneous. However, this effective nonlinearity is limited by the finite response time of the medium and corrections are expected in the non-adiabatic regime. Actually, the relaxation of the nonlinearity is known to have a deep influence on the electronic wave packet dynamics~\cite{kenkre1,kenkre2,campbell,Molina}. Using non-adiabatic nonlinear models,  V.M. Kenkre {\it et al}~\cite{kenkre1,kenkre2}  considered the problem of a two-level  system driven by a relaxation process of the nonlinearity. It was  shown that 
there is a stationary  self-trapping regime and a coexistence of static and dynamical transitions for certain degrees of nonlinearity and relaxation time~\cite{kenkre1}. These features contrast with the persistent oscillatory dynamics displayed by the adiabatic nonlinear dimer model in both delocalized and self-trapped regimes~\cite{campbell}. Moreover, it was shown that localized stationary states of a dimer model are destroyed above a critical temperature~\cite{kenkre2}. 

In this letter, we  address to the question regarding the influence of non-instantaneous nonlinearity on the long distance electronic transport in linear chains. This physical scenario is particularly suited to the study of electronic transport in macromolecules, such as DNA, on which the electron-phonon coupling is expected to play a relevant role~\cite{DNA}. Here, we will effectively take into account the non-adiabatic nature of the electron-phonon interaction by introducing a delayed third-order nonlinearity in the one-electron time-dependent Schr\"odinger equation. Delayed differential equations usually describe physical systems presenting a finite response time~\cite{referiram3}.
Within a tight-binding approach, the time-evolution of the coefficients of the wave vector expanded in the localized orbitals basis ($|\Psi(t)\rangle=\sum_jc_j(t)|j\rangle$) takes the form
\begin{equation}
i\dot{c}_j(t)=V(c_{j+1}(t)+ c_{j-1}(t)) -\chi|c_j(t-\tau)|^2c_j(t),
\label{DNLSE}
\end{equation}
where we used units of $\hbar=1$ and considered the on-site potentials $\epsilon_j=0$. $V$ is the hopping integral between nearest-neighbor sites, and $\chi$ is a nonlinear parameter which is proportional to the local electron-phonon coupling~\cite{Molina}.  The non-adiabatic nature of the  electron-phonon  interaction is represented by a delayed contribution to the on-site potential proportional to  $|c_j(t-\tau)|^2$ where $\tau$ is the typical response time. To analyze the wave packet propagation, we
solve Eq.(\ref{DNLSE}) using the fourth-order Runge-Kutta method
to  obtain the temporal evolution of an initially localized wave packet ($|\Psi(t=0)\rangle=\sum_jc_j(0)|j\rangle$ with $c_j(0)=\delta_{j,j_0}$). We follow the time evolution of  the amplitude of the wave function at the initial site,
calculating the so called return
probability~\cite{Nazareno99},
\begin{equation}\label{return}
R_0(t) \equiv |c_{j_0}(t)|^2 ~.
\end{equation}
Usually, the electron escapes from its initial position
when the amplitude $c_{j_0}(t)$ vanishes as $t$ evolves.
Conversely, the amplitude remains finite for a self-trapped
wave packet. In order to have a better description of the wave packet  dynamics,
we also compute the time-dependent participation function defined as 
\begin{equation}
P(t)=1/\sum_{j}|c_j(t)|^4~.
\label{participation}
\end{equation}
The participation function $P(t)$ gives an estimate of the number of sites over which the wave packet is spread at time $t$. In the long-time regime, its scaling behavior can also be used to distinguish between localized and delocalized wave packets~\cite{Nazareno99}. In particular, the asymptotic participation number becomes size-independent for localized wave packets. On the other hand, it scales linearly with the chain size in the delocalized regime.

\begin{figure}[t!]
\includegraphics*[width=4.1cm,clip]{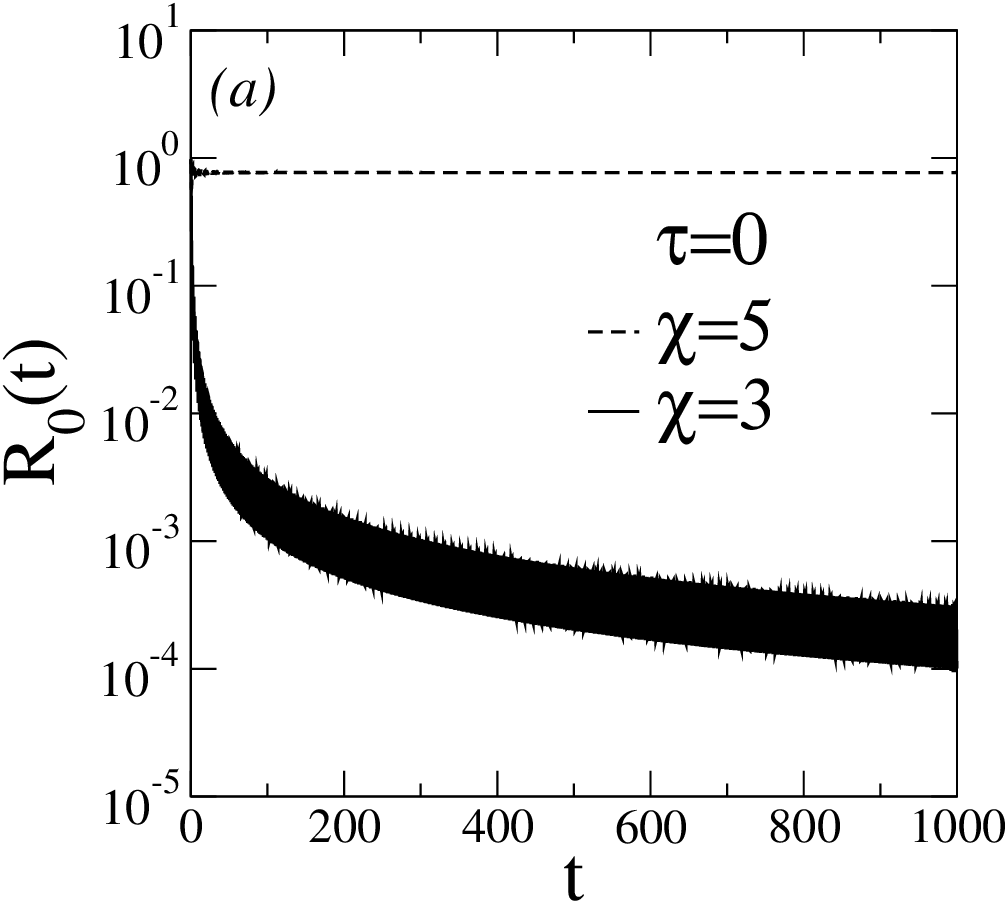} 
\includegraphics*[width=4.1cm,clip]{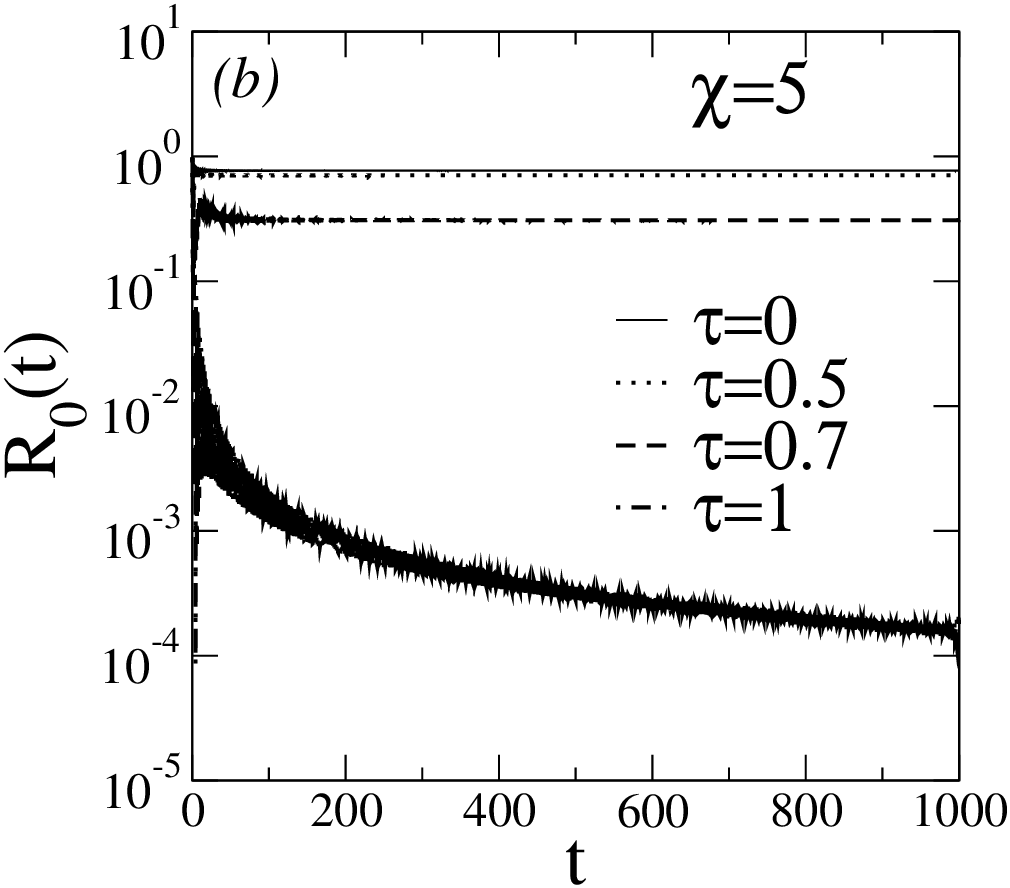}\\
\includegraphics*[width=4.1cm,clip]{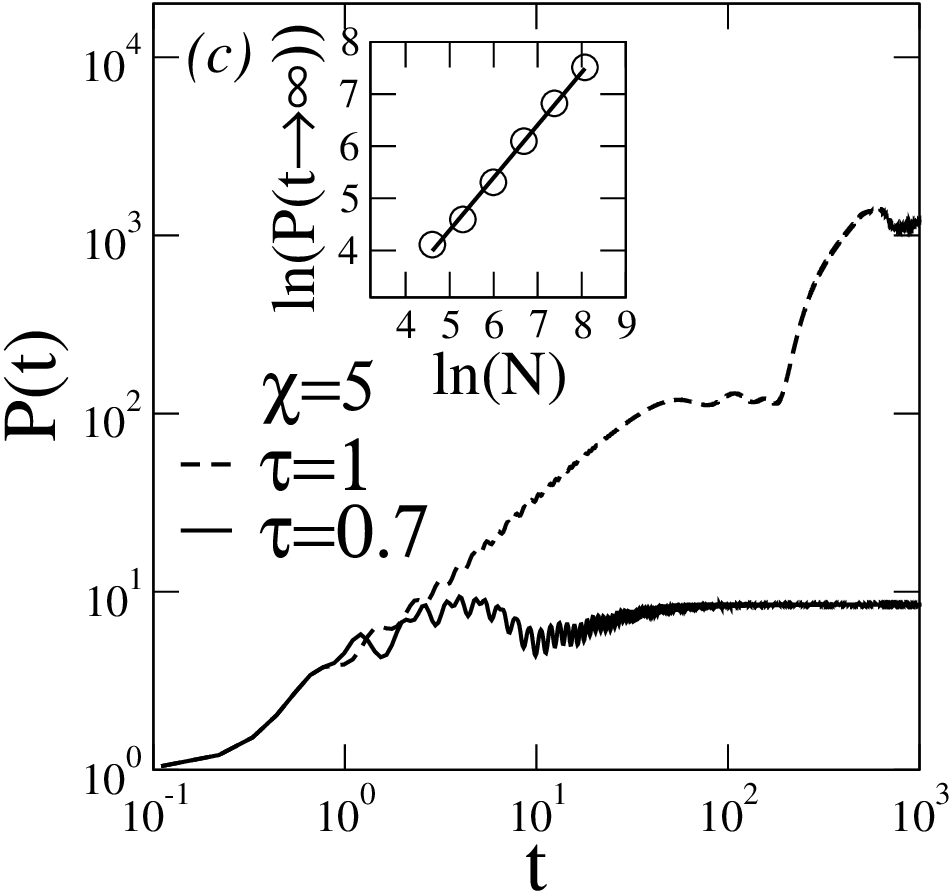}
\includegraphics*[width=4.1cm,clip]{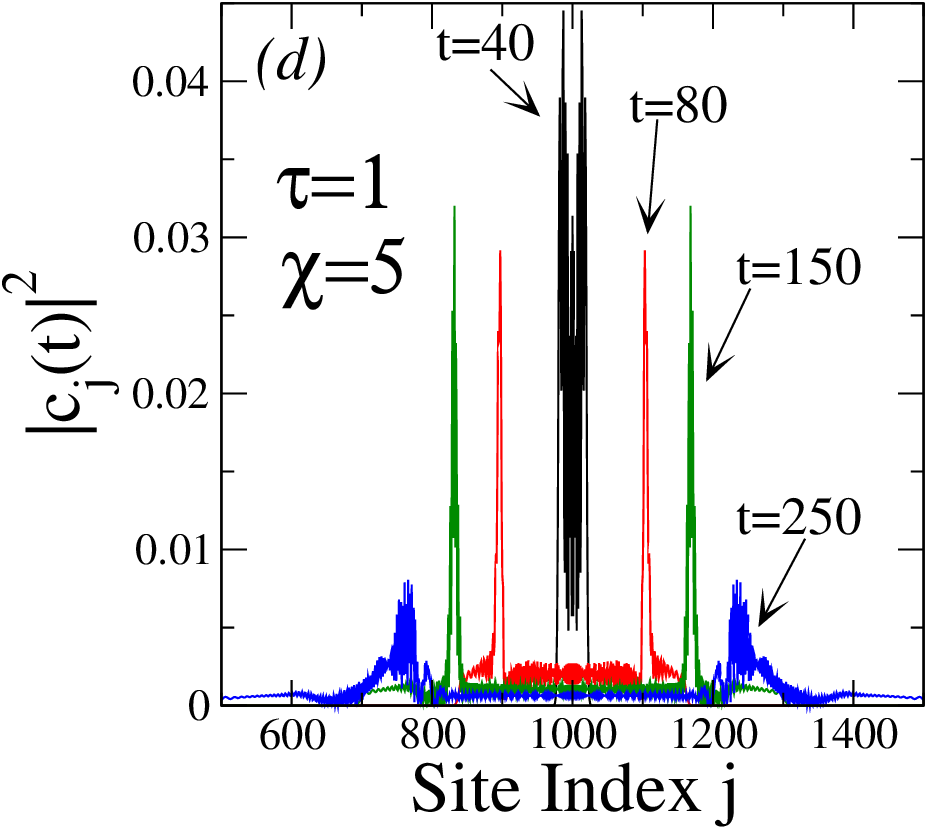}
\caption{(a) Return probability $R_0(t)$ for $\chi=3$ and $\chi=5$ in a chain with $N=2\times 10^3$ sites with an instantaneous nonlinearity. $R_0(t)$ asymptotes a constant value  for $\chi>3.5$, corresponding to a self-trapped wave packet. It continuously decay for $\chi <3.5$ as the wave packet becomes delocalized. (b) $R_0(t)$ for $\chi=5$ with different values of $\tau$.  Even for such strong nonlinearity, $R_0(t\rightarrow\infty)$ becomes of the order of $1/N$ for $\tau=1$. This breakdown of  self-trapping is a delay induced transition. (c) The participation number $P(t)$ for $\tau=0.7$ and $\tau=1$ for $\chi=5$. A long delay time allows for the spread of an initially localized wave packet. The inset shows the linear finite-size scaling of the asymptotic participation number for $\tau=1$, typical of delocalized states. (d)(Color online) Snapshots of the wave packet in the initial transient regime ($t=10$), within the plateau ($t=80, 150$) on which the width of the wave packet around the peaks is roughly constant in time, and after the plateau ($t=250$) showing the smearing out of the wave-front. 
} 
\label{fig1}
\end{figure}

In our numerical simulations, we set the parameter $V$ as the energy unit. Further, we considered the
open boundary condition with the wave packet initially located at the chain center. However, the main following analysis 
regarding the self-trapped/delocalized transition and phase diagram holds for the periodic  boundary condition as well. 
We assume $c_j(t<0)=0$ for any site $j$. This condition corresponds to the situation on which the electron is injected 
in the system at time $t=0$, with the electronic density being null for negative times. 
In fig.~\ref{fig1}(a), we report the time-evolution of the return probability on a chain with 
$N=2\times 10^3$ sites for the case of instantaneous nonlinearity ($\tau=0$) for distinct strengths 
of the nonlinear parameter $\chi$. The plots show the usual behavior exhibited  by the  DNLSE with
 adiabatic electron-phonon interaction~\cite{Xiong1}. For strong nonlinearities $\chi > 3.5,$ 
the return probability $R_0(t)$ approaches to a constant value in the a long-time limit, thus indicating that 
the wave packet is localized~\cite{Johansson,kundu}.  
This behavior is the so called self-trapping, on which the electron remains localized around its initial 
position due to its coupling with the lattice vibrations. The critical nonlinearity is of
the order of the energy bandwidth because $\chi$ accounts for the effective on-site potential felt by the
initially localized electron. If $\chi$ is much larger than the energy bandwidth, this state can not be efficiently
mixed with the other energy states.  For $\chi < 3.5$, 
 the asymptotic return probability is vanishingly small (of the order of $1/N$), 
which characterizes the regime of extended states. 
In Fig.~\ref{fig1}(b), we focus on the effect of the nonlinearity delay time $\tau$ on the self-trapping transition.  
We consider an electron-phonon coupling sufficiently strong to induce  self-trapping in the adiabatic limit. 
 We observe  that the gradual increase in the delay time leads to a decrease of the return probability. 
 This delocalization process, corresponding to a breakdown of  self-trapping, is a delay-induced transition. 
In Fig.~\ref{fig1}(c), we show data for the time dependent participation number $P(t)$ in this strong  nonlinear 
regime ($\chi=5$).  The self-trapping phenomenon is indeed
destroyed when the delay time is increased, with the asymptotic participation number becoming of the order of 
the chain size. It is interesting to notice that the participation number exhibits an intermediate plateau in 
the delocalized regime. The time evolution of $P(t)$ in this regime reflects the own wave packet dynamics [see Fig.~\ref{fig1}(d)]. 
During the initial stage, the wave packet splits from its single peak structure to a structure with two peaks that move
in opposite direction~\cite{katsanos}. The participation number grows almost linearly during this initial process. After this transient, the
wave-fronts display a soliton-like behavior, with the width around the peaks being roughly time-independent~\cite{Johansson}.
In this regime, the participation number becomes time-independent because the main contribution for it comes from the
region around the wave packet peaks. Further, there is a dynamical transition to a regime on which the wave-front peaks lose
their soliton-like behavior. The width of the wave-front start to increase in time and, consequently, the participation number
revivals its growth. It is interesting to stress that a dynamical transition has also been found in nonlinear dimers with non-adiabatic
nonlinearity~\cite{kenkre1}. 
Finally, the participation number achieves its final saturation when the wave-fronts reach the chain boundaries.       
The linear finite-size scaling, shown in the inset, clearly characterizes the delocalized nature of the asymptotic 
wave packet in this regime. 

\begin{figure}[t!]
\includegraphics*[width=4.7cm,clip]{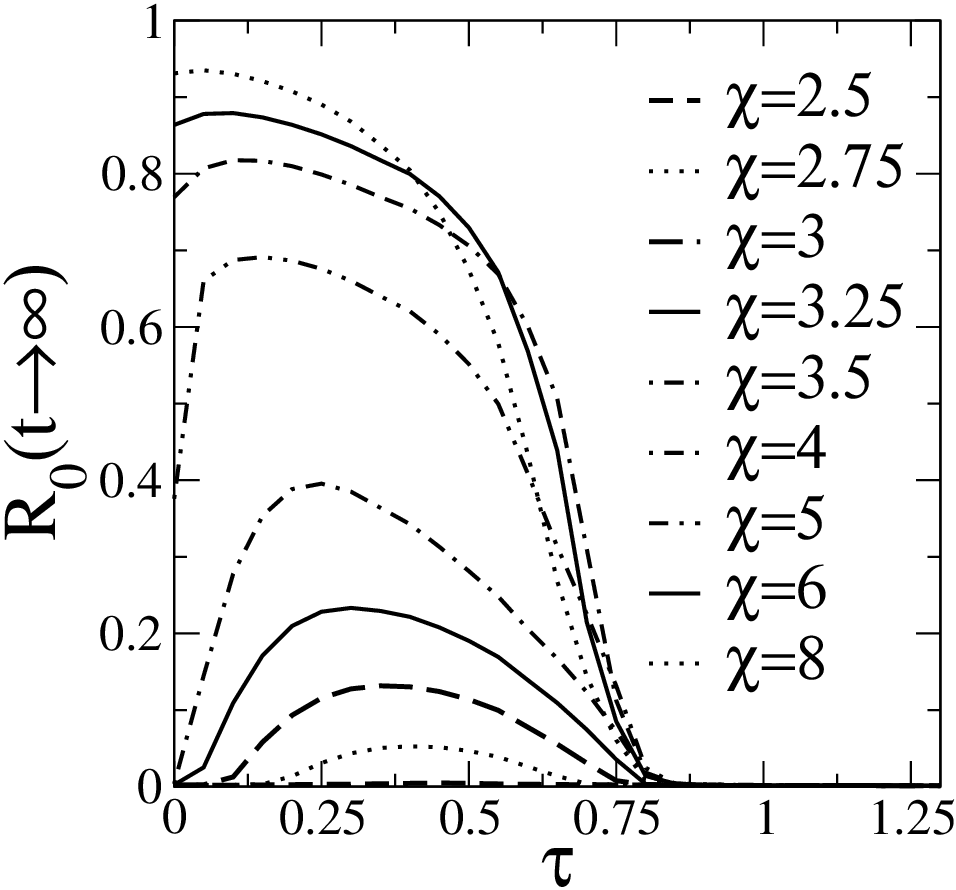} 
\caption{$R_0(t\rightarrow \infty)$ versus the delay time  $\tau$ with $\chi=2.5$ up to $8$.  A breakdown of the self-trapping phenomenon takes place for large delay times. For intermediate electron-phonon interactions $2.5<\chi<3.5$, we observe that a short delay can induce self-trapping [$R_0(t\rightarrow \infty )>0$] which is absent in the limit of instantaneous nonlinearity. In this range of nonlinear couplings, a sequence of delocalized$\rightarrow$self-trapped$\rightarrow$delocalized  regimes is depicted as the delay time is increased. 
} 
\label{fig2}
\end{figure}

Data for the long-time return probability 
$R_0(t\rightarrow \infty)$ versus the delay time $\tau$ are shown in Fig.~\ref{fig2} 
for nonlinearity strengths ranging from $\chi=2.5$ up to $8$. 
The reported values are averages in the statistically stationary regime of the return probability.
These values are finite and independent of the chain size when self-trapping takes place. On the other hand, the
asymptotic return probability vanishes as $1/N$ in the delocalized regime. 
For $\chi<2.5$, the asymptotic return probability approaches to zero irrespective to the delay time.  
In contrast with 
the delocalization transition induced by large delay times, a short delay can promote self-trapping 
[$R_0(t\rightarrow \infty )>0$] for nonlinearities slightly below the adiabatic critical one $2.5<\chi<3.5$. The 
physical reason for the decreasing of the critical nonlinear strength in the regime of fast nonlinear responses
is that the nonlinearity comes from the electron coupling with the electronic density
 at a previous time. With the electron 
still relatively localized around its initial position in 
the beginning of the wave packet time-evolution, 
the retarded electronic density at the initial site is larger
 than the instantaneous one. 
Therefore, a smaller nonlinear strength is required to promote the self-trapping in 
comparison with the instantaneous case.  
Therefore, our results indicate that the self-trapping regime has a reentrant character taking place in a finite
range of short delay times. 
\begin{figure}[t!]
\includegraphics*[width=4.7cm,clip]{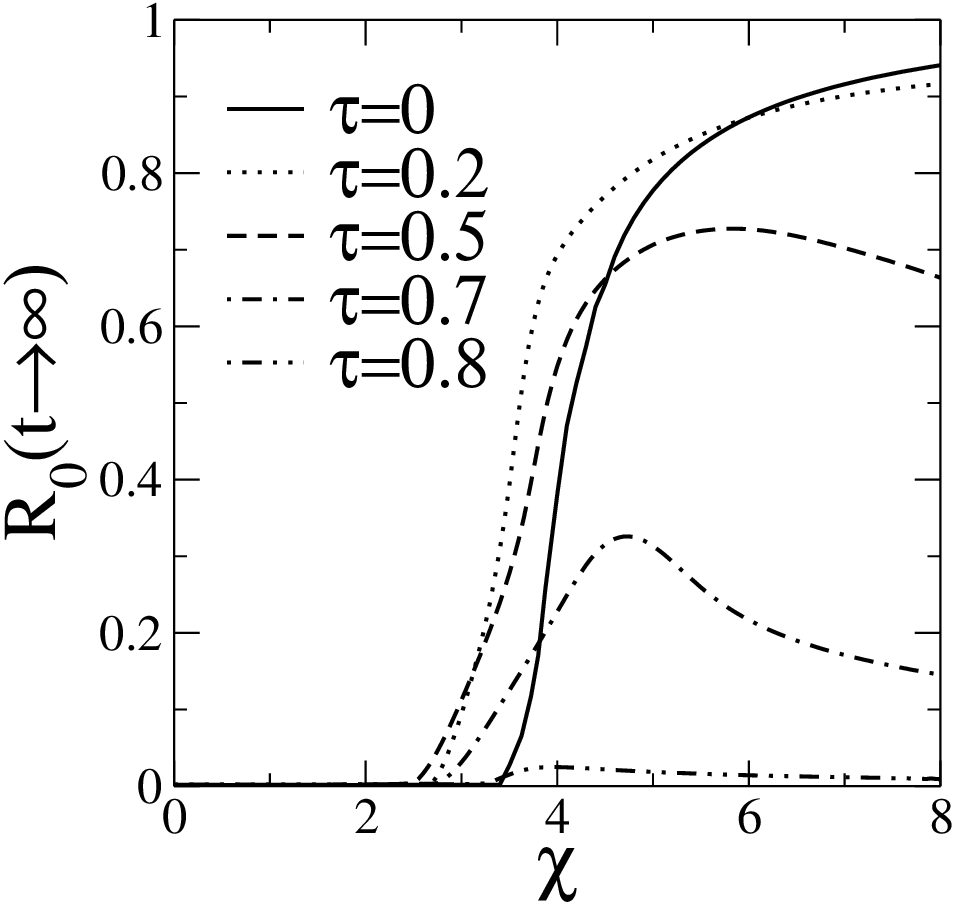} 
\caption{$R_0(t\rightarrow \infty)$ versus the strength of the electron-phonon coupling $\chi$ for $\tau=0$ up to $0.8$. A delayed nonlinearity can induce  self-trapping even for nonlinear couplings below the critical value in the adiabatic limit $\chi=3.5$.
} 
\label{fig3}
\end{figure}
\begin{figure}[t!]
\includegraphics*[width=4.7cm,clip]{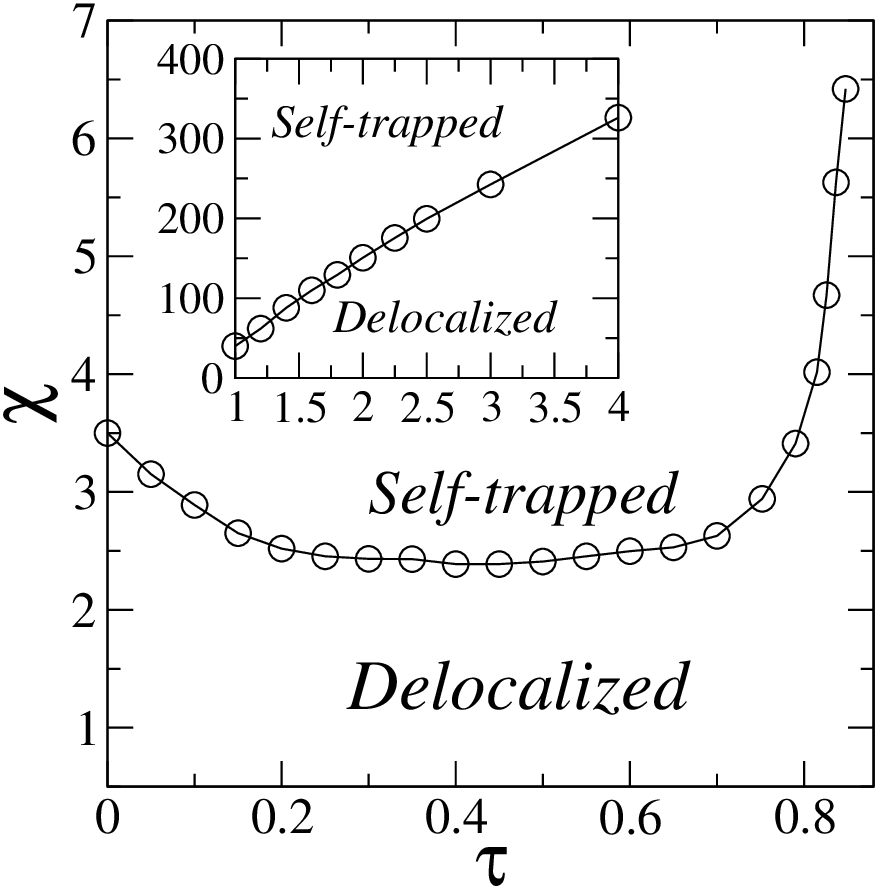} 
\caption{Phase diagram in the $\chi \times \tau$ parameter space showing the delocalized/self-trapped transition induced by a time delayed electron-phonon interaction. The regime of delocalized wave packets was characterized using the 	
criterion $R_0(t\rightarrow \infty)\approx 1/N$ and $P(t\rightarrow \infty) \propto N$. The critical nonlinearity has a non-monotonic behavior in the regime of short response times. In the inset, we show the critical line in the regime of longer response times at which much stronger nonlinearities are required to promote self-trapping.
} 
\label{fig4}
\end{figure}
In Fig.~\ref{fig3} we plot some numerical results for $R_0(t\rightarrow \infty)$ versus the strength of the electron-phonon coupling $\chi$. Here, we considered delay times ranging from $\tau=0$ up to $0.8$. These results give further support to the above described picture: (i) for large delays the self-trapping phenomenon disappears (see results for $\tau=0.8$) in the range of nonlinearity strengths shown; (ii) for small but finite delays, self-trapping starts  below the critical nonlinearity of the adiabatic limit. It is interesting to notice that the return probability reaches a maximum at a finite nonlinear coupling in the regime of short delay times. This non-monotonic behavior also contrasts with the continuous increase of the return probability for the case of an instantaneous nonlinearity. Finally,   we
plot the phase diagram in the $\chi \times \tau$ plane in Fig.~\ref{fig4}. The reentrant character of the transition is reflected by the non-monotonic dependence of $\chi_c$ as the delay time is increased. The extended states were characterized using the 	
criterion $R_0(t\rightarrow \infty)\approx 1/N$ and $P(t\rightarrow \infty) \propto N$. The slight decrease of the critical nonlinear strength for short delay times is replaced by a fast increase at longer delays, as shown in the inset.  Actually, the wave packet spreads substantially before the nonlinearity takes action in slowly responding media. As a consequence, a strong nonlinearity is required to stop
 the subsequent wave packet spread. 

In summary, we considered the problem of one electron moving on a chain under the influence of a delayed 
third-order nonlinearity which effectively incorporates the effect a non-adiabatic  
electron-phonon interaction. In the case of an instantaneous 
nonlinear response, this system presents a delocalized/self-trapped transition at a critical nonlinear 
strength $\chi\simeq 3.5$~\cite{Johansson,Xiong1}. We found that for the case of long delay times, 
self-trapping takes place only for much stronger nonlinearities. As a consequence, the electronic transport 
retains its metallic character in slowly responding media for a wide range of nonlinear strengths.  
On the other hand, we report 
a reentrant behavior of the self-trapping transition point in the regime of fast nonlinear responses with 
self-trapping occurring even below the critical nonlinear strength of 
instantaneously responding media. In a finite range of nonlinearities below the adiabatic critical value, 
a short delay time can promote the stabilization of self-trapping. A second transition takes place at a 
longer delay time above which the wave packet becomes again delocalized. 

It is important to stress that the presently reported delay-induced self-trapping/delocalization transition is of a nature distinct from the one taking place in dimer-like structures~\cite{kenkre1}. While in the later the delay promotes the decay of the oscillations in the population levels and a transition from two distinct oscillatory dynamical regimes, the presently reported transition deals with the long-distance electron transport. The here reported phenomenon shall be inherent to dynamical systems described by delayed nonlinear differential equations. Therefore the influence of the nonlinear response time on the electronic wave packet dynamics is expected to have a similar counterpart in general (acoustic, electromagnetic) wave transport in nonlinear media. It would be interesting to have further studies along this direction.

This work was financially supported by CNPq-Rede Nanobioestruturas, CAPES, FINEP
(Brazilian research agencies) and FAPEAL (Alagoas State agency).


\begin{thebibliography}{50}
\bibitem{Johansson}M. Johansson, M. H\"ornquist, and R. Riklund, Phys. Rev. B {\bf 52}, 231 (1995).
\bibitem{kundu}P.K. Datta and K. Kundu, Phys. Rev. B {\bf 53}, 14929 (1996).
\bibitem{Xiong1}Z. Pan, S. Xiong, and C. Gong, Phys. Rev. E {\bf 56} 4744 (1997).
\bibitem{ivanchenko}M.V. Ivanchenko, Phys. Rev. Lett. {\bf 102}, 175507 (2009).
\bibitem{flach} S. Flach, D.O. Krimer, and Ch. Skokos, Phys. Rev. Lett. {\bf 102}, 024101 (2009).
\bibitem{skokos} Ch. Skokos, D.O. Krimer, S. Komineas, and S. Flach, Phys. Rev. E {\bf 79}, 056211 (2009).
\bibitem{capone}G. Sangiovanni, M. Capone, C. Castellani, and M. Grilli, Phys. Rev. Lett. {\bf 94}, 026401 (2005).
\bibitem{prl1}G. Kopidakis, S. Komineas, S. Flach, and S. Aubry,  Phys. Rev. Lett. {\bf 100}, 084103  (2008).
\bibitem{prl2}A.S. Pikovsky and D.L. Shepelyansky,  Phys. Rev. Lett. {\bf 100}, 094101  (2008).
\bibitem{prl3}D. Hajnal and R. Schilling,  Phys. Rev. Lett. {\bf 101}, 124101  (2008).
\bibitem{prl4}Y. Lahini, A. Avidan, F. Pozzi, M. Sorel, R. Morandotti, D.N. Christodoulides, and Y. Silberberg,  Phys. Rev. Lett. {\bf 100}, 013906  (2008).
\bibitem{sea}A. Maluckov, Lj. Hadzievski, N. Lazarides, and G.P. Tsironis,  Phys. Rev. E {\bf 79}, 025601(R) (2009).
\bibitem{kenkre1}V.M. Kenkre and H.-L. Wu,  Phys. Rev. B {\bf 39},6907  (1989).
\bibitem{kenkre2}P. Grigolini, H.-L. Wu, and V.M. Kenkre, Phys. Rev. B {\bf 40},7045  (1989). 
\bibitem{campbell}V.M. Kenkre and D.K. Campbell, Phys. Rev. B {\bf 34}, 4959 (1986).
\bibitem{Molina}D. Chen, M.I. Molina, and G.P. Tsironis, J. Phys.: Condens. Matter {\bf 5}, 8689 (1993) .
\bibitem{DNA}Mikrajuddin, K. Okuyama, and F.G. Shi, Phys. Rev. B {\bf 61}, 8224 (2000); W. Zhang, A.O. Govorov, and S.E. Ulloa, Phys. Rev. B {\bf 66}, 060303(R) (2002); M. Taniguchi and T. Kawai, Phys. Rev. E {\bf 72}, 061909 (2005); E. Macia, Phys. Rev. B {\bf 76}, 245123 (2007).
\bibitem{referiram3}M. Berezowski, Chaos Soliton Fractals {\bf 12},  83 (2001); J.M. Buld\'u, J. Garc\'{\i}a-Ojalvo, and M.C. Torrent, Phys. Rev. E {\bf 69}, 046207 (2004); R. Vicente, S. Tang, J. Mulet, C.R. Mirasso, and J.-M. Liu, Phys. Rev E {\bf 73}, 047201 (2006); F. Ishida and Y.E. Sawada, Phys. Rev. E {\bf 75}, 012901 (2007).
\bibitem{Nazareno99}H.N. Nazareno and P.E. de Brito, Phys. Rev. B {\bf 60}, 4629 (1999).
\bibitem{katsanos}D.E. Katsanos, S.N. Evangelou, and S.J. Xiong, Phys. Rev. B {\bf 51}, 895 (1995).












\end{thebibliography}
\end{document}